\documentclass[structabstract]{aa} 
\usepackage{txfonts}
\usepackage{natbib}
\usepackage{graphicx}
\usepackage{url}
\usepackage[]{longtable}
\begin{document}
 \newcommand{\ma}{\mbox{m\AA}}
\newcommand{\ms }{\mbox{m s$^{-1} ~$}}
\newcommand{\cms }{\mbox{cm s$^{-1} ~$}}
\newcommand{\kms }{\mbox{km s$^{-1}~ $}}
 \newcommand{\cm}{cm$^{-2}$}
 \title{A  Frequency Comb calibrated Solar Atlas  \thanks{Observations based on the ESO 3.6m Telescope at La Silla, Chile}}
\author{
P. \, Molaro
\inst{1} 
\and M. Esposito\inst{3}
\and S.  \,  Monai \inst{1}
\and G.  \,  Lo Curto  \inst{2} 
\and J. I. Gonz\'alez Hern\'andez   \inst{3} 
\and T. W. H\"ansch  \inst{4} 
\and R.   Holzwarth  \inst{4,5} 
\and  A.   Manescau  \inst{2} 
\and  L. Pasquini.  \inst{2} 
\and R.  A. Probst  \inst{4} 
\and R. Rebolo  \inst{3} 
\and  T.  Steinmetz  \inst{5} 
\and  Th. Udem  \inst{4} 
\and T.      Wilken    \inst{4}   }
\offprints{P. Molaro,   \email { molaro@oats.inaf.it} }
\institute  {$^1$ INAF Osservatorio Astronomico di Trieste, Via G.\,B.\,Tiepolo 11,
34143 Trieste, Italy\\
$^2$  ESO, Karl-Schwarzschild-Strasse 2  85748 Garching, Germany\\
$^3$  IAC,  C/ Via Lactea, E38205,  La Laguna  (Tenerife), Spain\\
$^4$  Max-Planck-Institut f\" ur Quantenoptik, Hans-Kopfermann-Str. 1, 85748 Garching, Germany.\\
$^5$  Menlo Systems GmbH, Am Klopferspitz 19a, 82152 Martinsried, Germany\\
\email{molaro@oats.inaf.it}
}
\date{Accepted: 15 Oct 2013 }
\abstract
{The solar spectrum is a primary reference    for the study of  physical processes in  stars  and their variation during  activity cycles.  
 }
{ In November 2010 an experiment  with a prototype of a Laser Frequency  Comb (LFC) calibration system was performed with the HARPS spectrograph of the 3.6m ESO telescope at La Silla during which  high signal-to-noise  spectra of   the Moon  were obtained. We   exploit  those Echelle spectra  to study a portion of the optical integrated solar spectrum    and in particular  to  determine    the  solar photospheric  line positions.  }
{The DAOSPEC program  is used to measure solar line positions through gaussian fitting in an automatic way.  The solar spectra are  calibrated both with an LFC and  a Th-Ar.}
{    We first apply the LFC solar spectrum  to characterize the CCDs of the HARPS spectrograph. The comparison of the LFC and Th-Ar calibrated spectra reveals  S-type distortions  on  each order along the whole spectral range with an amplitude of $\pm$  40 \ms. This   confirms the pattern  found by Wilken et al. (2010) on a single order and extends the detection of the distortions  to the whole  analyzed region revealing that the precise shape  varies with  wavelength.  A new data reduction is implemented to deal with 
CCD pixel inequalities  to obtain a wavelength corrected solar spectrum.  By using this spectrum   we provide a   new  LFC calibrated solar  atlas   with   400   line positions in the range of $476-530$,  and 175 lines in the $ 534-585   $ nm range   corresponding to the   LFC bandwidth.    The  new LFC atlas  is  consistent on average  with that based  on FTS solar spectra,    but  it  improves      the accuracy  of individual lines  by a significant factor  reaching  a mean   value  of   $\approx$ 10  \ms .}
{ The LFC--based   solar line wavelengths are  essentially  free of major instrumental effects  and  provide a  reference   for  absolute  solar line positions  at the date of Nov 2010,  i.e.  an epoch of  low   solar activity.  We suggest that future  LFC  observations could  be used  to trace  small   radial velocity  changes  of the whole solar photospheric spectrum  in connection with  the solar cycle  and for direct comparison with the predicted line positions of 3D radiative hydrodynamical models of the solar photosphere.   The LFC  calibrated solar atlas can be also used to verify the accuracy of  ground or space spectrographs    by means of the  solar spectrum. 
}


\maketitle

\begin{figure}[]
\begin{center} 
\resizebox{!}{6.0cm}{\includegraphics[clip=true,angle=0]{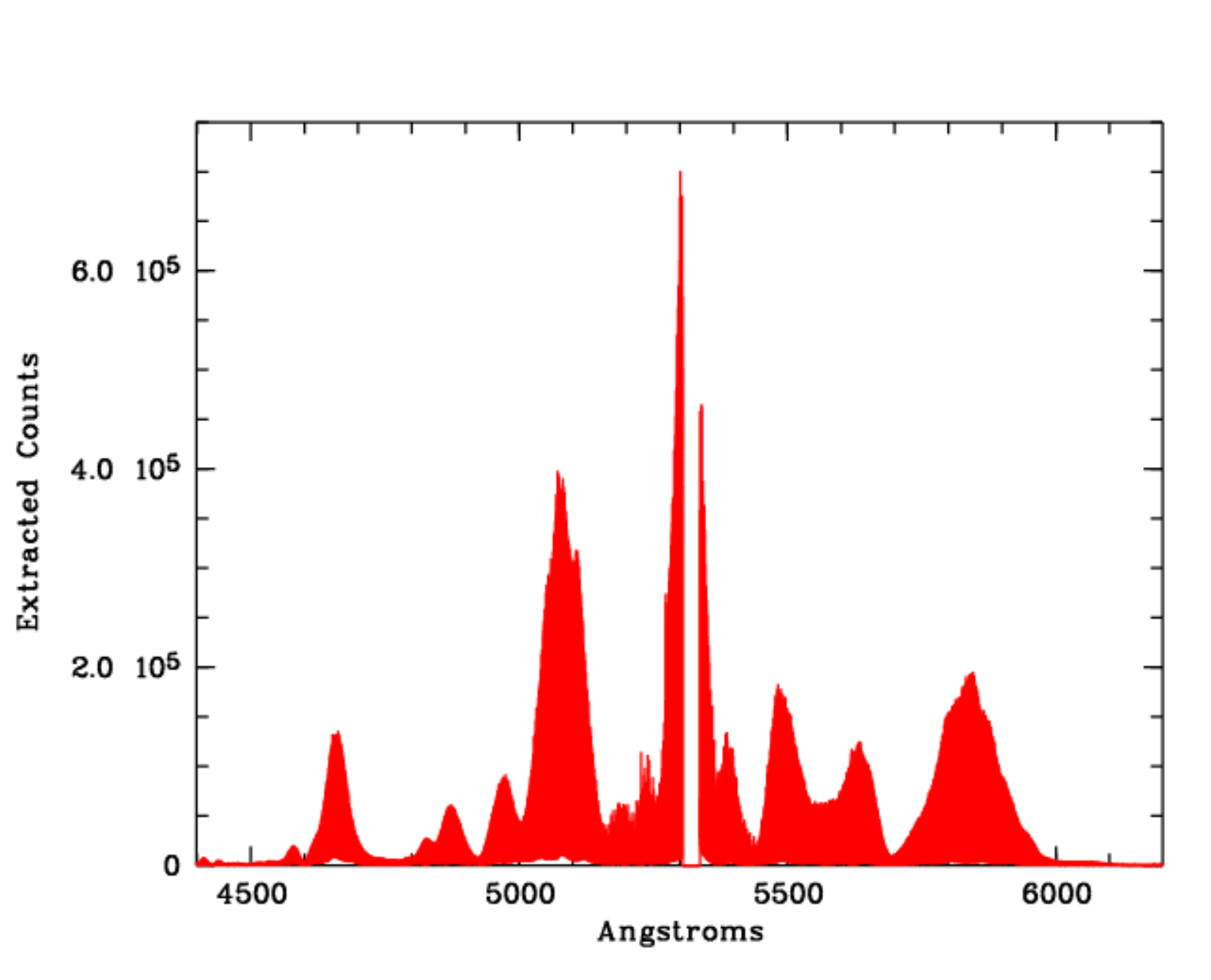}}
\caption{  LFC spectrum in HARPS fiber A.  The  gap  between  530 and  534 nm is the lost region between the two detectors. The emission lines of the LFC  are not resolved on this scale. A zoom is shown in Fig.2.  } 
\end{center}
\label{fig1}
\end{figure}  
\section{Introduction}

\begin{figure}[]
\begin{center} 
\resizebox{!}{6.0cm}{\includegraphics[clip=true,angle=0]{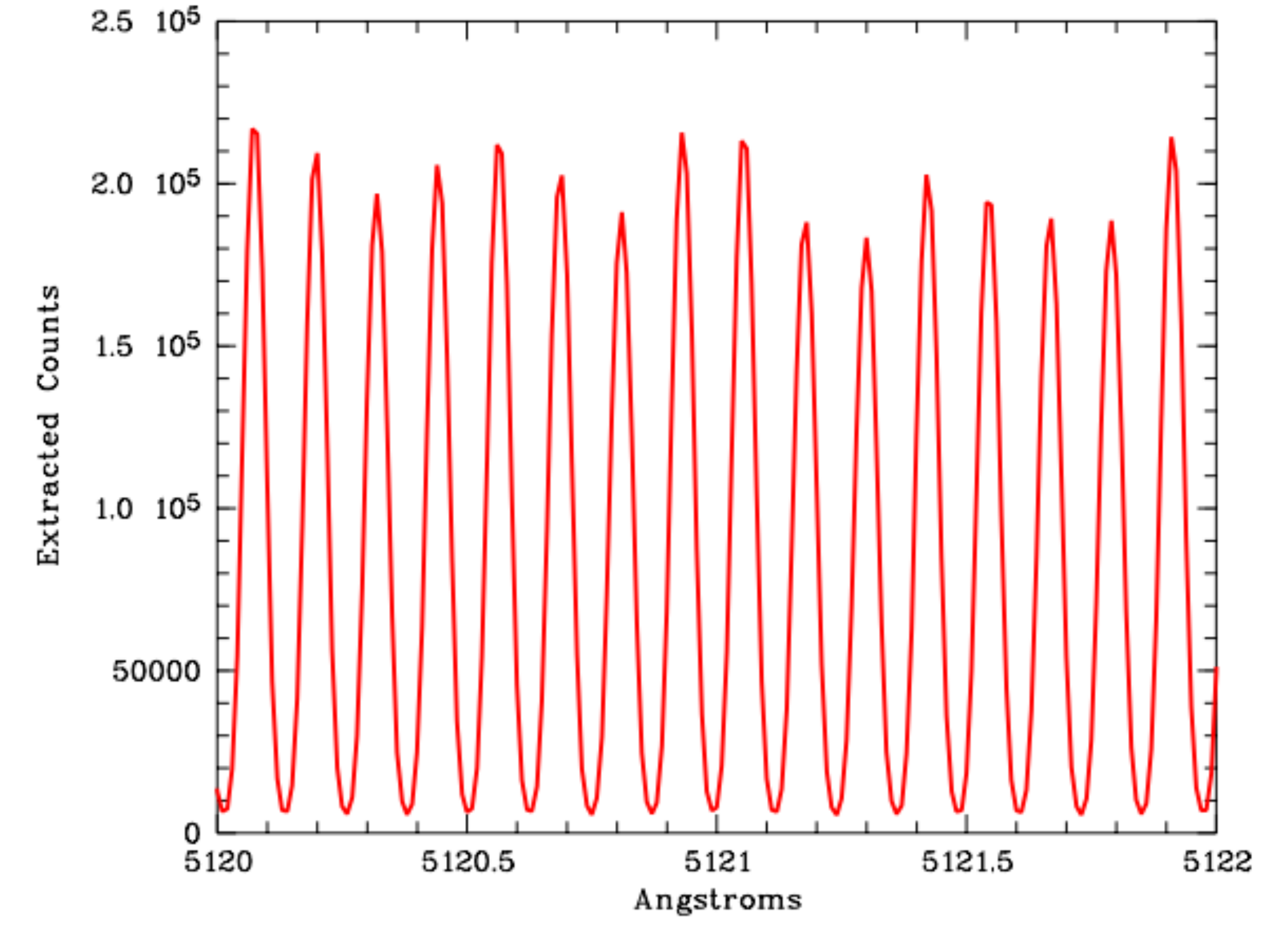}}
\caption{ Zoom of   LFC spectrum in the 512.0-512.2  nm~ region revealing a line density of 80 lines per nm .} 
\end{center}
\label{fig2}
\end{figure}

The optical and infrared  spectrum  most   used as  reference for   solar and stellar studies is the one obtained with the
 Fourier Transform Spectrometer (FTS) at  the McMath-Pierce Solar Telescope at Kitt Peak  National Observatory    \citep{kurucz84, kurucz06}. 
The  spectrum  covers the spectral region  from 296 to 1300 nm  and achieves a signal-to-noise ratio of  $\approx$ 2500 with  a resolving power of R$\approx$ 400 000.      However, the FTS solar spectrum is a  merger of    several observations taken at different positions of the solar disk   performed from November 1980 to June 1981 that  were  later combined  to  produce an integrated  solar flux.   In addition, the   wavelength's precision of the FTS composite solar spectrum degrades with the distance of the  sub-spectrum  from the FTS zero point. As a result,     the observed wavelengths  of the solar lines   are poorly determined, with  an  uncertainty  at the level of  a  hundred   \ms  at  best \citep{AllendePrieto98a, AllendePrieto98b}.  Disk-integrated solar spectra were taken with the
un-focused sunlight   NSO FTS spectra \citep{gra97, wal11}.
\citet{mol11} and \citet{mol12} used  HARPS spectra of  asteroids  to  improve  the precision of the line position  relative to those provided by FTS spectra.
  
The Doppler shifts   of  the photospheric lines are  related to the gravitational redshift and to dominant motions of the convective layers in which  they form. When integrated over the disk, the latter    are averaged over  a variety of inhomogeneities across the stellar surface that produce the granulation and  that  may also vary in time in connection to the solar magnetic activity  
\citep{lin03}.   
The  spectrograph  used to record the solar spectrum can also play a non-negligible role in introducing small distortions in wavelength.  The spectral accuracy   of  conventional cross-dispersed spectrographs  ultimately depends on the accuracy of the wavelength calibration. In the
visible this is usually achieved by  using arc lamp spectra such as those
produced in hollow cathode Thorium lamps \citep{meg57, gia70}.
Thorium has  thousands of  spectral features in the visible  domain,  it is mono-isotopic  and  has no 
hyperfine structure,  and produces   narrow and highly symmetric line
profiles. For all these reasons,  Th-Ar hollow cathode
lamps have become the standard for wavelength calibration of
astronomical spectrographs.
The reference for the  Th wavelengths is
the Atlas of the Thorium Spectrum obtained with the McMath-Pierce Fourier Transform Spectrometer (FTS) of the National
Solar Observatory at Kitt Peak \citep{PE83}, (PE83), with  a spectral resolution of $\approx$  600 000. The atlas gives the position of $\approx$  11 500 lines between 300 and 1100 nm with a
quoted wavenumber accuracy   at 550.0 nm  ranging from 16  to  82 \ms .
 However, most of  of them  are 
blends  of individual components and with a spectrograph with resolving power of 100 000 the number of usable
lines  decreases  to $\approx$  4000. \citet{lovis07}  taking advantage of the stability of the High Accuracy Radial velocity Planet Searcher (HARPS) spectrograph suggested how to increase the number of usable lines, in particular of faint lines and how to improve  their wavelength position. However, as we  show in this paper, these positions are affected by the HARPS detector system and their application should remain confined to this spectrograph.  
In the last years a new wavelength calibration technique based on a Laser Frequency Comb  (LFC)  able to produce a sequence of  equally spaced emission lines with similar intensity has been suggested as an ideal calibrator \citep{mur07} and subsequently demonstrated by  \citet{wil12,yca12,phi12}.

    With the first LFC experiment  performed with HARPS at the 3.6~m ESO telescope at La Silla, Chile, spectral   distortions    were found  at the level of  about $\pm$ 40 \ms  in order  N. 120  \citep[ reproduced here in Fig. 6]{wil10a}. To  achieve large CCDs, manufacturers  employ  a method  which consists in  butting a master lithographic
block of 1024x512 pixels. Each of the two HARPS CCDs is  build up with  16 of these master blocks. At
the boundaries of the  blocks  the pixel size is not exactly  15 $\mu$m and so  is the distance between the center of the  two adjacent active pixels on the two sides of the discontinuity. \citet{wil10a}  suggested that   manufacturing imperfection of pixel sizes in the HARPS CCDs might be responsible for an adaptation of the dispersion curves in the  wavelength calibration analysis  that produces the observed 
S-type distortion.     These distortions should be present in all   HARPS spectra even though  the spectrograph is able to deliver sub \ms  relative precision    and should   also be present in    other Echelle spectrographs  which make use of CCDs manufactured with the  stitching technique across the dispersion direction.
In the present paper we analyze HARPS solar spectra as reflected by   the Moon and  calibrated with an  LFC spanning more than  about 100 nm.   The aim is to provide a high-precision  solar atlas  with  reduced  instrumental uncertainties   and to characterize the HARPS spectrograph over a range wider   than what  tested by the pioneering paper of \citet{wil10a}.

%
\begin{table}
\caption{\footnotesize Journal of observations. The corresponding number of solar spots is also given as an indicator of the solar activity.}       
 \label{table:1}
\centering          
\begin{tabular}{ c c c r rrrr}     
\hline\hline       
 Date  & UT start exp  &N. S. Spots & Exp.(sec.)  &       &   &   \\ 
              &   &    & &&&          &       \\   
\hline                    

 2010-11-25& 05-47-20&  14.4   & 60 &  &  &  &   \\
2010-11-25 & 05-49-42&  14.4   & 150 &  &  &  &   \\
2010-11-25  & 05-53-19&  14.4   & 60 &  &  &  &   \\
  2010-11-25& 05-55-20&  14.4   & 120 &  &  &  &   \\
2010-11-25 & 05-57-54&  14.4   & 120 &  &  &  &   \\
\hline
\end{tabular}
\end{table}
%

%
\begin{table*}
\caption{\footnotesize DAOSPEC analysis of different Moon solar spectra. RV$_{c}$ is the expected radial velocity  computed with  JPL ephemerides.   RV$_{o}^b$ ,VR$_{o}^r$  and $\sigma$  are the mean radial velocities and    dispersion of all the  identified lines by DAOSPEC with respect to the input line list,  separately for the two CCDs. $\Delta RV(o-c)^{b}$ and $\Delta RV(o-c)^{r}$ are the difference between the two quantities.    
  }        
\label{table:2}
\centering          
\begin{tabular}{ccccccccccc}     
\hline\hline       
Observation &     RV$_{c}$ & RV$_{o}^{b}$   &  $\sigma$  & N.  lines &  $\Delta RV(o-c)^{b}$ & RV$_{o}^{r}$       & $\sigma$         & N. lines  & $\Delta RV(o-c)^{r}$& \\ 
       &     \kms      &  \kms      &    \kms   & & \kms & \kms &&\ms &\kms & \\   
\hline                    
2010-11-25, 05-47-20& -1.190& -1.181& 0.267 &494 &  0.009 & -1.169 & 0.061 & 239 &   0.020 & \\
2010-11-25, 05-49-42& -1.187&-1.180   & 0.248 &484 &  0.006 &  -1.166 & 0.057 & 240 &   0.020 & \\
2010-11-25, 05-53-19& -1.182&  -1.176   & 0.256 &487 &  0.007 & -1.158  & 0.059 & 236 &   0.024 & \\
2010-11-25, 05-55-20  & -1.179&  -1.176  & 0.263 &490 & 0.004 &  -1.157 & 0.056 & 234 &   0.022 & \\
2010-11-25, 05-57-54& -1.176&  -1.171   & 0.264 &493 &  0.005&  -1.158  & 0.057 & 236 &   0.019 & \\
 
 \hline
\end{tabular}
\end{table*}

\section{Observations and data reduction}

\subsection{The Astro-Comb  }

 Since  2007   ESO, in collaboration with  Menlo Systems and the Max-Planck-Institut f\"ur Quantenoptik (MPQ),  has developed an LFC prototype, called also astro-comb,  dedicated to astronomical observations.  
 Four campaigns   with HARPS
 at the 3.6 m  telescope in La Silla  were organized to
test the global performance \citep{wil10a,wil10b, wil12}.

A  LFC consists in  a multitude of equally spaced optical frequencies over a bandwidth of  several tens to hundreds of THz. Its light source is a passively mode-locked laser emitting a train of femtosecond pulses. In the frequency domain, a set of very sharp spectral lines at frequencies $f_n = n\cdot f_r + f_0$ is obtained. Here, $n$ is an integer, $f_r$  is the repetition rate of the optical pulses, and $f_0$ is the so-called offset frequency. Both $f_r$ and $f_0$ are radio-frequencies  that are stabilized to an atomic clock which directly transfers the accuracy and stability of the atomic clock into the optical frequency domain.

The first LFC tested on an astronomical spectrograph was in the infrared \citep{stei08}. LFCs are the ideal calibrators for astronomical high resolution spectrographs if they can cover the spectral bandwidth of the spectrograph with a sufficiently flat spectrum and if their line spacing is adapted to the spectrograph  resolution. Typically, this requires a  line spacing of 10 to 30 GHz.

The ESO astro-comb is based on fibre-laser technology  which greatly enhances robustness and reliability, and enables the use of high-power amplifiers. Yb-doped fibre-lasers, as used in this comb, are particularly suited since the second harmonic of the central wavelength at about 1030 nm is in the centre of the visible range. However, the required fibre length limits the round-trip time in the oscillator, and thus its repetition rate. Commercially available fibre-based LFCs therefore typically have a line-spacing of 250 MHz. For astronomical applications, the line-spacing is increased by using Fabry-Perot cavities as a high-resolution spectral filter, suppressing the unwanted lines. The LFC for the HARPS spectrograph, as employed here and described in \citet{wil12}, was filtered to a line spacing of 18 GHz by 3 concatenated cavities. This was followed by frequency doubling the LFC spectrum, yielding a center wavelength of about 525 nm, and subsequent spectral broadening in a tapered photonic crystal fibre. 

The stability of the LFC system was tested by comparing 1400 spectra collected between November 2010 and January 2011 \citep{wil12} . These spectra showed that on short time scales the level at which the data no longer follow the photon noise is $\approx$ 2.5 \cms for the LFC, while it is at  $\approx$ 10 \cms   for ThAr. By using the LFC calibrated spectra of HD 75289, \citet{wil12}  could also reconstruct the orbit of a Jupiter size exoplanet with a period of 3.5 days known to orbit around this solar-type star.

\begin{figure}[]
\begin{center} 
\resizebox{!}{6.0cm}{\includegraphics[clip=true,angle=0]{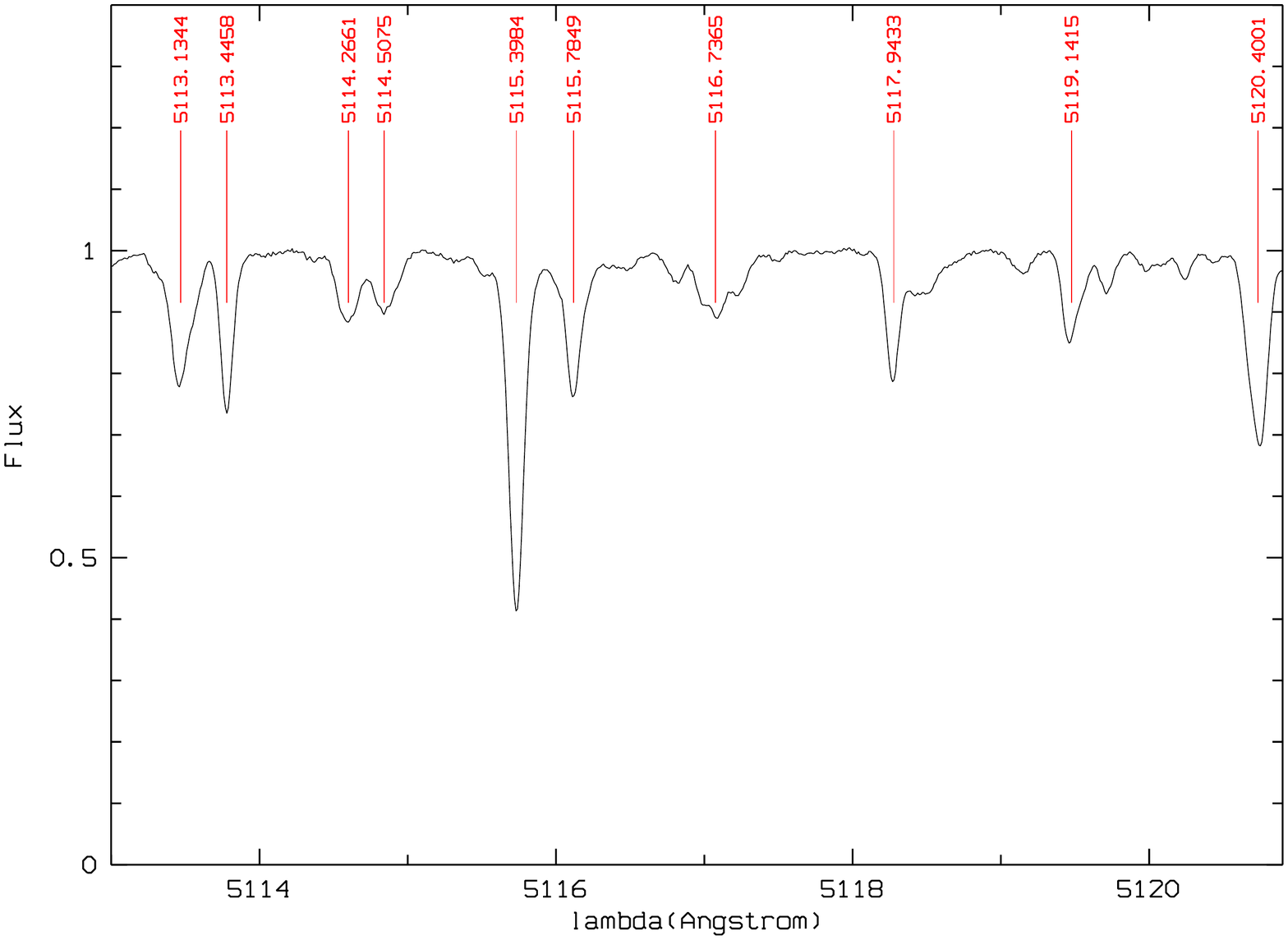}}
\caption{ A portion of the solar spectrum  in which the lines used for position measurements  are marked.} 
\end{center}
\label{fig3}
\end{figure}

\subsection{The observations}

   The  observations of the Moon  analyzed in this paper were made  with the HARPS spectrograph during  Nov 2010 observing run. The journal of the observations is   reported in Table \ref{table:1} together with some basic information.  The observations were performed pointing towards the Moon's center, and the Moon's illuminated fraction was of 86\%.
  HARPS spectra   are  taken at   a resolving power of R = $ \lambda / \delta \lambda$ = 115 000  covering   the spectral  range from $380$ to $690$ nm with     a small  gap at $530-539$ nm. The   Point Spread Function  in the order centers   was  sampled with 4.1 pixels   15 $\mu$m  wide.  The fibers feeding  HARPS  are   equipped 
with a  double  scrambler to provide both slit and  pupil stability. This is an  important difference compared to slit spectrographs   where non-uniform  slit illumination induces radial velocity differences between different exposures.  HARPS   is able to  deliver  a sequence of observations over  a 500-day baseline with a dispersion of 64 \cms  along the exoplanet radial velocity orbital change.
 
 The LFC spectrum  is imaged onto parts of both of HARPS' CCDs for a total   wavelength range   of  $\approx$  125 nm, spanning from 465 to 590 nm.
The whole LFC spectrum   is shown unresolved in  Fig 1. Since  the flux  has considerable variations in  intensity over the spectrum,   a good calibration can be obtained in a somewhat more restricted range from about 476 and 585 nm with a gap between the two detectors at 530 to 534 nm. 
 Note that the structure in the LFC spectral envelope was a feature of 2010  test while   present LFC are generated with a very constant spectral envelope.
The LFC spectrum has  a very high   line density  and individual lines are not resolved on this scale. A  zoom  into a  spectral  window  of   0.2   nm ~   at 512.0 nm~ is reproduced  in Fig.  2   showing the uniform and non-blended  
line pattern with a line density of 80 lines per nm.
 The LFC has    $\approx$  350 lines per order, while   
the Th-Ar lamp has  less than  $\approx$  100 lines per order with different intensity, sometimes blended or saturated. 
The LFC spectra were used both to obtain the wavelength calibration and to correct for instrumental drifts.
The LFC emission lines are identified and fitted using gaussian profiles. The wavelength solution {is obtained by means of a 
third degree polynomial along the CCD. The average number of modes per order in the wavelength range we are studying is 
of  350, corresponding to a mode separation of 18 GHz. While fiber A of HARPS is fed with the moonlight from the telescope, the LFC is launched into fiber B, providing a simultaneous reference spectrum to monitor  overnight instrumental  drifts. The instrumental drift is computed from the displacement of the LFC spectrum acquired simultaneously with 
the  Sun  spectra on the adjacent fiber. For each LFC mode the displacement is weighted by the signal 
collected on that mode.


 \section{Measuring  solar line positions}

We used   the    DAOSPEC program    to measure  the position of the photospheric  lines on     solar   spectra  calibrated with the LFC.
DAOSPEC is a program  developed by P. Stetson that  automatically finds absorption lines in a stellar spectrum, fits the continuum and identifies lines with the help of a laboratory line list \citep{ste08}.    Spectral lines are modeled by  gaussian profiles and identify blends through an estimation of the FWHM adjusted on the actual measurement of the line widths.  As such DAOSPEC is  not adequate to measure strongly saturated lines which are on the square-root or damping parts of the curve of growth or  to reveal  the line asymmetries, namely the $\it C$  shape of the line bisector of  photospheric lines.  It is well known that the blueward hook of the $\it C$ originates from the upflowing granules and dominates the bisector close to the continuum while the intergranular components are more gaussian and slightly redshifted.  The result is that strictly speaking the radial velocity of a line is a function of the line depth.   DAOSPEC   provides the line position under the gaussian approximation  and  computes the  mean  radial velocity    from  the  lines of  the  input lines list  identified in the spectrum.   As such this  is an approximation but has the advantage to be easily reproduced and measured.  An example of the automatic identifications made by DAOSPEC in a small portion of the solar spectrum is shown in  Fig. 3.  For the initial line list we have adopted the one derived  by  \citet{mol12} in their HARPS solar data, which contains  many  strong and  unblended  lines in the spectral region covered by the LFC. We set   a  threshold of 0.0025  nm   for the minimum equivalent width (EW)   to avoid problems with relatively weak lines  and performed the  analysis on the two different HARPS  CCDs separately.  The initial line list is a list of the observed lines in the sun thus already incorporating  the convective blueshift and the gravitational redshift.
		 The mean radial velocities of the    spectra computed from all identified lines by DAOSPEC with respect to the input line list are provided  in  Table \ref{table:2} separately for the blue (RV$_{o}^b$) and red (RV$_{o}^r$)  CCDs .

 \begin{figure*}[]
\begin{center} 
\resizebox{!}{11.cm}{\includegraphics[clip=true,angle=0]{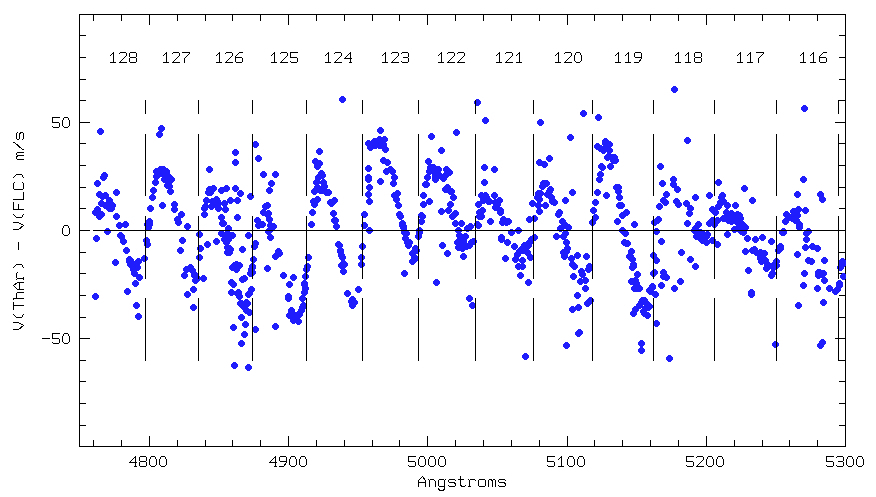}}
\caption{  Velocity difference between the lines  as measured from the solar spectrum calibrated with Th-Ar and the same spectrum calibrated with the LFC: blue portion. The vertical dashed lines and the number mark the Echelle orders } 
\end{center}
\label{fig4}
\end{figure*}

\begin{figure*}[]
\begin{center} 
\resizebox{!}{11.0cm}{\includegraphics[clip=true,angle=0]{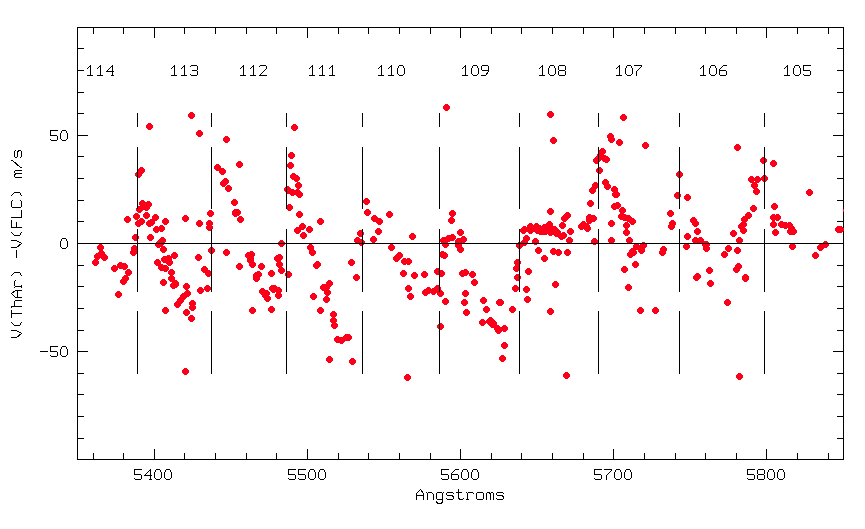}}
\caption{ Velocity difference between the lines  as measured from the solar spectrum calibrated with Th-Ar and the same spectrum calibrated with the LFC: red portion.
The vertical dashed lines and the number mark the Echelle orders. } 
\end{center}
\label{fig5}
\end{figure*}

 The theoretical expected radial velocity of the solar spectrum accounts for the motions of the Moon with respect to the observer and  with respect to the Sun. For the latter we also estimate  a component of  3.2 \ms due to  the rotation of the Moon, since the sunlight hits the receding  lunar hemisphere before being reflected with an angle of 43$^o$.  
   These  are  computed by using the JPL ephemerides and their sum is  provided in the second  column of Table \ref{table:2}. 
  The mean difference  between  the measured radial velocity RV$_{o}^{b}$ or RV$_{o}^{r}$    and   the expected value RV$_{c}$ as computed with  JPL ephemerides    is  $\Delta RV(o-c)^{r}$  $\approx$  +21.7 $\pm 2.4$  \ms in the red and  $\Delta RV(o-c)^{b}$  = 5.7 $\pm 2.2$  \ms in the blue.
  Any difference with the predicted velocity  can be due either to an intrinsic shift of the reference solar spectrum   induced by  solar activity or to an error in the wavelengths of  the solar spectrum lines taken as a reference.  Since in  2006,  the year of the reference spectrum,  the Sun   showed   six solar spots, slightly less  than the 14 of the epoch of the present observations,   we assume that  
  the radial velocity differences    reflect uncertainties more  likely coming from the reference spectrum used as an input list, i.e. the subset of the line list provided by \citet{mol12} which was   generated from HARPS spectra of CERES  calibrated with a conventional Th-Ar technique. 
 The standard deviation of the distribution of the deviations  of the line position with respect to derived  radial velocity  returned by the DAOSPEC analysis is  of  $\approx$ 260   \ms  and  of  $\approx$ 60 \ms in the blue and  red, respectively.  In the following  section  we will show that a significant contribution to the scatter of the red CCD  originates in the instrumental distortions introduced by CCD manufacturing defects. However, the blue region shows a considerably  larger dispersion which is more likely originated in the solar spectrum used as a reference.

\begin{table}
\caption[]{LFC Solar Atlas in the 476.0-585.0 nm region.  Wavelengths  ($\lambda$ ) are in air transformed  from vacuum assuming a refractivity of air for standard composition, at a temperature T=15 $^o C$ and an atmospheric pressure of  P= 760 Torr. The value reported are   the mean of the  positions measured  in the five   spectra together with  their standard deviation. The
full Table is available on the online electronic version. }
\label{table:3}
\begin{tabular}{lrrc}
\hline
\hline
~~~~~~~$\lambda$    & $\sigma$~~~~~  &EW  & Ident  \\ 
~~~~~~~\AA    &  \ms  & m\AA &  \\ 
\hline
5593.73963 &51.73  &41.4 &NI1 \\
5594.65538 &41.16  &64.0 &FE1 \\
5598.30493 &30.29  &82.1 &FE1 \\
5600.03586 &31.28  &28.5 &MN2 \\
5600.22349 &16.07  &36.7 &FE1 \\
5601.28320 &40.22  &97.7 &CA1 \\
5614.78097 &18.89  &41.8 &NI1 \\
5618.63788 & 6.06  &49.9 &FE1 \\
5619.60420 &13.80  &33.3 &FE1 \\
5620.47933 &19.18  &36.8 &FE1\\
5624.02998 &11.99  &49.7 &FE  \\
5625.32288 &27.45  &39.5 &NI1 \\
5633.95031 &51.04  &65.5 &FE1 \\
5635.82653 &16.82  &34.8 &FE1 \\
5637.12063 &15.95  &33.4 &NI1 \\
5637.40814 &13.14  &41.7 &OS1 \\
5638.26528 & 6.53  &73.9 &FE1 \\
5641.44472 & 3.12  &63.0 &FE1 \\
5644.12610 &22.26  &35.0 &FE2 \\
5645.61161 & 3.76  &32.0 &SI1 \\
5649.67584 & 8.73  &31.5 &NI1 \\
5649.99071 &12.45  &34.1 &FE1 \\
5650.68848 & 9.56  &35.9 &FE1 \\
5652.32118 &11.43  &25.2 &FE1 \\
           \hline   
\end{tabular}       
\end{table}


\section{Detection of HARPS spectral distortions}
  
 As an application of the newly derived solar line list we show how it  can be used to reveal instrumental distortions  of HARPS  spectra  calibrated with Th-Ar lamps. For this purpose    we compared the line positions measured in the   spectrum  calibrated in wavelength with the Th-Ar with those measured in the same spectrum but calibrated   with the  LFC.   The differences between the two sets of measurements    on the full set of 1245 solar lines with EW greater than 0.0025 nm detected in the range are shown in  Fig.  4 for the blue and   Fig. 5 for the red.  The difference in wavelengths  reveals a characteristic pattern which is remarkably similar in each order from one order to the other. The pattern  is a kind of S-type with positive velocity differences at the beginning of each order. It is only in the last 3 orders of the red portion investigated that the pattern is less clear or even absent. On average the mean radial velocity in the whole red range is of -0.8  $\pm$ 22 \ms and -0.16 $\pm$ 21.6 \ms in the blue range,  but could be quite offset from zero in small spectral ranges. For instance, in order N.109 the mean value  is of -9.2 $\pm$ 22.5 \ms while in order N. 123 it is of 14.2 $\pm$ 19.1 \ms.
  In general,  the pattern is very similar to the one revealed by \citet{wil10a} in the first  experiment with an  LFC  on  HARPS. Owing to the high line density of
the LFC spectra,  \citet{wil10a} were able to characterize the detectors by revealing    the effect of the
stitching pattern of the CCD fabrication on the wavelength calibration.  The S-type shape is the result of the  adaptation of the  polynomial solution to the limited number of Th lines if non identical pixel sizes are present. The presence of non identical pixels in proximity of the CCD stitch goes unnoticed when Th-Ar calibration
is used, because  of  the  low number of lines present in the orders, but produces 
 a peak-to-peak radial velocity distortion which  reaches  $\approx$ 
  80 \ms  in the  Echelle  order 120. 

A zoom of  the results for the  Echelle order 120   is   shown in  Fig. 6   together with  what found by \citet{wil10a} for the same order,   which  shows  a good overlap.  There are only  few points deviating  significantly from the S-type shape that are due to relatively weak lines where the error in the  measurement of the line position is particularly large. This comparison shows that  although the number density of useful solar lines is considerably lower  than  the LFC emission,  they are considerably more numerous   than the Th lines and sufficient to reveal   the  wavelength shifts introduced by  CCD defects. 

Another way to see the same distortion is by measuring the difference in line position of the LFC spectrum once calibrated with the  Th-Ar solution, thus taking advantage of the  denser  LFC lines with respect to the solar spectrum.  Fig. 7 and  Fig. 8 show the velocity difference between the line positions  in  the LFC spectrum  calibrated with  the Th-Ar lamp and the same LFC spectrum  calibrated  on itself. The velocities derived with the Th-Ar calibration are corrected by -9 and -20 \ms since as  discussed in the previous section the Th-Ar provides slightly redshifted wavelengths in this portion of the spectrum. The orders are not merged and reveal an additional distortion at the edges.  However, a close  resemblance with the distortions revealed by using the solar lines can be seen. 
 In particular the deviation from zero of individual orders and the less  defined  S-type behaviour around 525.0 nm or in the red CCD are also confirmed. We have verified that the distortions remain stable over the period of these observations. Although we do not expect these to change, they should  be verified with further similar LFC experiments performed with a  longer time   baseline. The close similarity between the two results provides  confidence on the accuracy of the wavelengths of the LFC solar atlas. Therefore,  this atlas can be used  to  calibrate or to test the accuracy of any ground or space spectrograph from an acquisition of a solar spectrum and the comparison with the present atlas.

The observed inter-order  distortions are likely  induced by the  stitching pattern which is present in the HARPS CCDs.  Different  pixel sizes can be revealed by accurate flat fielding and in particular by technical LED  flats which illuminate the whole CCD. Summing up  of columns   of the LED  flats for the two HARPS CCDs  normal to the dispersion direction are shown in  Fig. 9.   The   technical flats   reveal  abrupt changes in the flux  involving 1 or 2 pixels in correspondence of the  CCD stitching but with an offset of 4 pixels, i.e. every 4+ n 512 pixels.   The changes in flux are  highly variable going from a minimum of $\approx$ 0.4 \% of the adjacent flux   at position 1540 up to $\approx$ 3 \% which is observed at position 1028. The  flux  changes reflect a change  in the sensitive area of the pixels close to the junctions.  Thus
 every 512 pixels there is  a discontinuity as the intra-pixel distance is slightly different along the stitching borders.   It is the combined effect  of these irregularities with the position and number of Th lines  in that order that produces  the observed spectral distortions, which are    similar  but not identical in all  spectral orders as revealed by our analysis  by means of the LFC calibrated solar atlas.


\begin{figure}[]
\begin{center} 
\resizebox{!}{6.5cm}{\includegraphics[clip=true,angle=0]{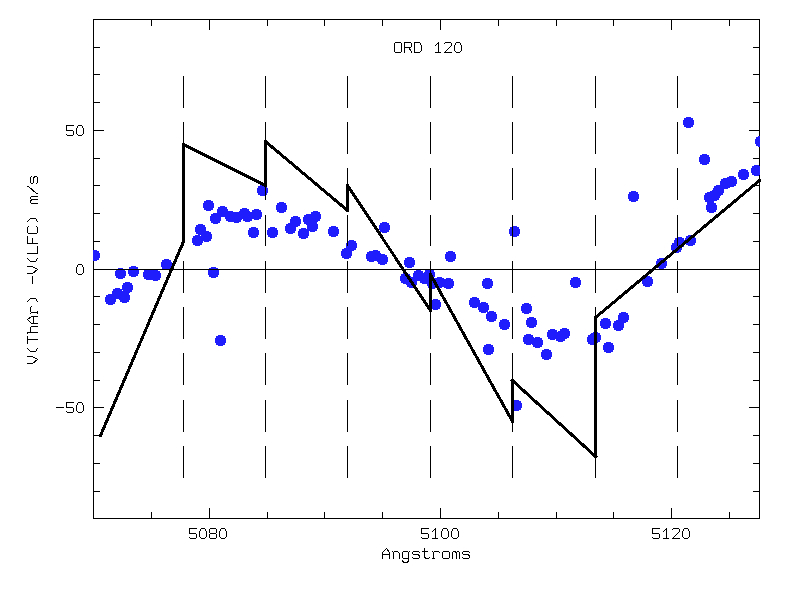}}
\caption{ As in Fig. 4 but zoomed on order  N. 120. This order was previously  studied by  \citet{wil10a} and their instrumental distortions are over-sketched as a black continuous line. The vertical dashed lines mark the 8 chunks of 512 pixels each.} 
\end{center}
\label{fig6}
\end{figure}







 \begin{figure}[]
\begin{center} 
\resizebox{!}{5.cm}{\includegraphics[clip=true,angle=0]{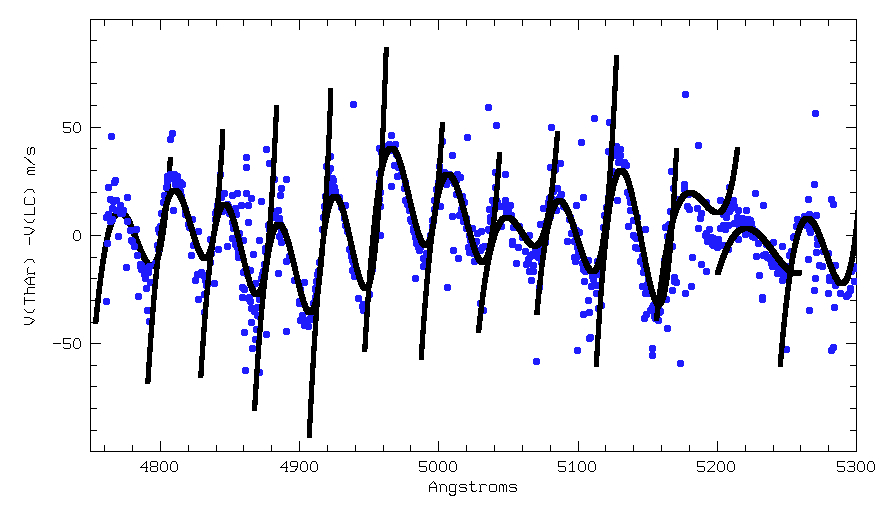}}
\caption{  Wavelength difference  between the lines  of the LFC spectrum when calibrated with Th-Ar: the blue CCD.  The difference here is performed in the non merged orders. The orders are easily identified since at the edges large deviations are observed. The points of Fig 4 based on the solar lines are over-plotted. } 
\end{center}
\label{fig7}
\end{figure}

 \begin{figure}[]
\begin{center} 
\resizebox{!}{5.cm}{\includegraphics[clip=true,angle=0]{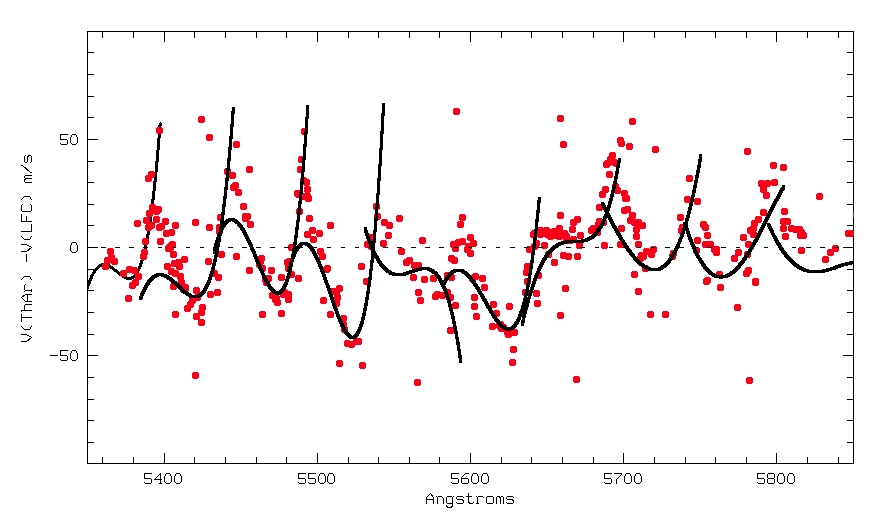}}
\caption{ As in Fig. 7 for the red CCD with over-plotted the data points of Fig. 5. } 
\end{center}
\label{fig8}
\end{figure}

\section{A LFC calibrated solar atlas}

We have improved the wavelength calibration in order to properly account for the pixel size discontinuities
occurring at the CCD stitches .  The spectral order were each divided
into 8 chunks of 512 pixel elements, as many as the number of master blocks along the
dispersion direction. For each chunk, the wavelength calibration
was obtained by fitting a third degree polynomial to the position
of the LFC modes. In each chunk we  had some 40-50 LFC lines 
to be used for the fit.  An example of the improvement in the residuals of the new wavelength calibration is shown in Fig. 10. The residuals  from  wavelength solutions restricted to eight portions of the CCD are significantly smaller and do not show spikes or trends as   when the CCD is considered as made up of pixels of equal size.
A similar approach cannot be followed with hollow cathode lamps 
 because of  paucity and  irregular distribution of the Th-Ar lines
lines.

Thus, thanks to the high spectral density of the LFC lines, and to their precision  we correct for CCD non-uniformities and obtained   a  solar  spectrum  which  is relatively free from major   instrumental effects affecting the wavelength calibration.  We used DAOSPEC to measure the solar line positions on these spectra.  
   Out of   the about 1200 lines initially selected,  we retained  only lines with   positions that did  not deviate significantly from the mean value measured in the  five  spectra of the Moon.   In this way the selected list is less vulnerable to  the presence of  blends and of  reduction problems. In fact, deviations larger than average   are   observed for lines falling  in correspondence of the wings of the strong lines of  H$_\alpha$, H$_\beta$ and of the NaI and MgI multiplets.    After a 3$\sigma$ clipping the final line list comprises  400 and 175   lines in the ranges $476.0-530.4$ and $534.0- 585.0$ nm, respectively.  The global radial velocity is adjusted to match the theoretical radial velocity expected from the Sun and therefore made independent from the accuracy of the radial velocity of the input list. The standard deviation of the difference in the velocities of  the five  solar spectra is also reported in   the second column of  Table \ref{table:3}. Average values  are of 12 and 15 \ms in the blue and red, respectively, and are   consistent with  the  photon noise error.    The table  provides  the first entries of the atlas  as an example of the content, while  
 the full table is available electronically . The  solar spectra adopted in this analysis and calibrated by means of the LFC are also made available in ascii  format for general use
 \footnote{Table 3 and the solar spectra  are only available in electronic form
at the CDS via anonymous ftp to cdsarc.u-strasbg.fr (130.79.128.5)
or via http://cdsweb.u-strasbg.fr/cgi-bin/qcat?J/A+A/}. The spectra are non-normalized spectra with a step size of 0.01 \AA~  and are separated   into a blue   (475.368 and 530.409) and a  red  (533.726 and 586.000) spectra   corresponding to the portion of spectra falling onto the two HARPS CCDs which have been calibrated  with the LFC. The spectra provided are  corrected by the  motion of the observer towards the Moon and by the motion of the Moon towards the Sun, the sum of which is  reported in the  second column of Table 2.  The calibrated LFC of individual orders, which gives in formation of the shape and behaviour of the spectral point spread function, is also provided.

\section{The  Lovis \& Pepe (2007) Th-Ar line list}

 The detected inter-order distortions in HARPS spectra produced by  different pixel sizes in the  CCDs  have relevant bearings on the Th-Ar line list provided by  \citet{lovis07} (LP07) which is based on HARPS Th-Ar observations.
The authors   made use of the higher sensitivity of a  grating spectrograph such as HARPS compared to a Fourier Transform Spectrometer  to reduce the random noise of the  \citet{PE83} wavelengths. LP07
used a large set of Th-Ar spectra taken with HARPS over one month  to improve the measurement precision of many Th-Ar lines of  PE83.  
They identified  lines  which in HARPS showed no significant  wavelength variations with time and obtained their  positions with typical   rms  of 5 -10 \ms. Thus, they  suggested to replace  the wavelengths of individual PE83 lines  according to their observed position. The new  LP07  Th-Ar atlas  contains $\approx$ 4000 previously unidentified lines and  more than doubles the
number of lines available for wavelength calibration. However, for what discussed in the previous section it is very likely that the  Th lines in HARPS were falling in a slightly different position if compared to PE83 as  the result of the spectral distortions introduced by CCD defects which we  revealed by means of the LFC wavelength calibration. 
A new thorium line list based on  new measurements from the   FTS at NIST has been made  available  very recently \citep{red13}.   When the new  lines are compared with those of LP07 a large scatter is  confirmed. 

LP07 noted that  only some lines where  changing  position with time. They attributed  this to  either the fact that  these  lines experience significant shifts with changing lamp pressure or current,  or that they are  actually a blend with varying relative intensities of the lines composing the blend.  Considering what discussed in this paper,   it is more likely that lines that changed  position were closer to the critical points generating the spectral distortions and therefore  were  subject to  larger instabilities.

\begin{figure}[]
\begin{center} 
\resizebox{!}{6.5cm}{\includegraphics[clip=true,angle=0]{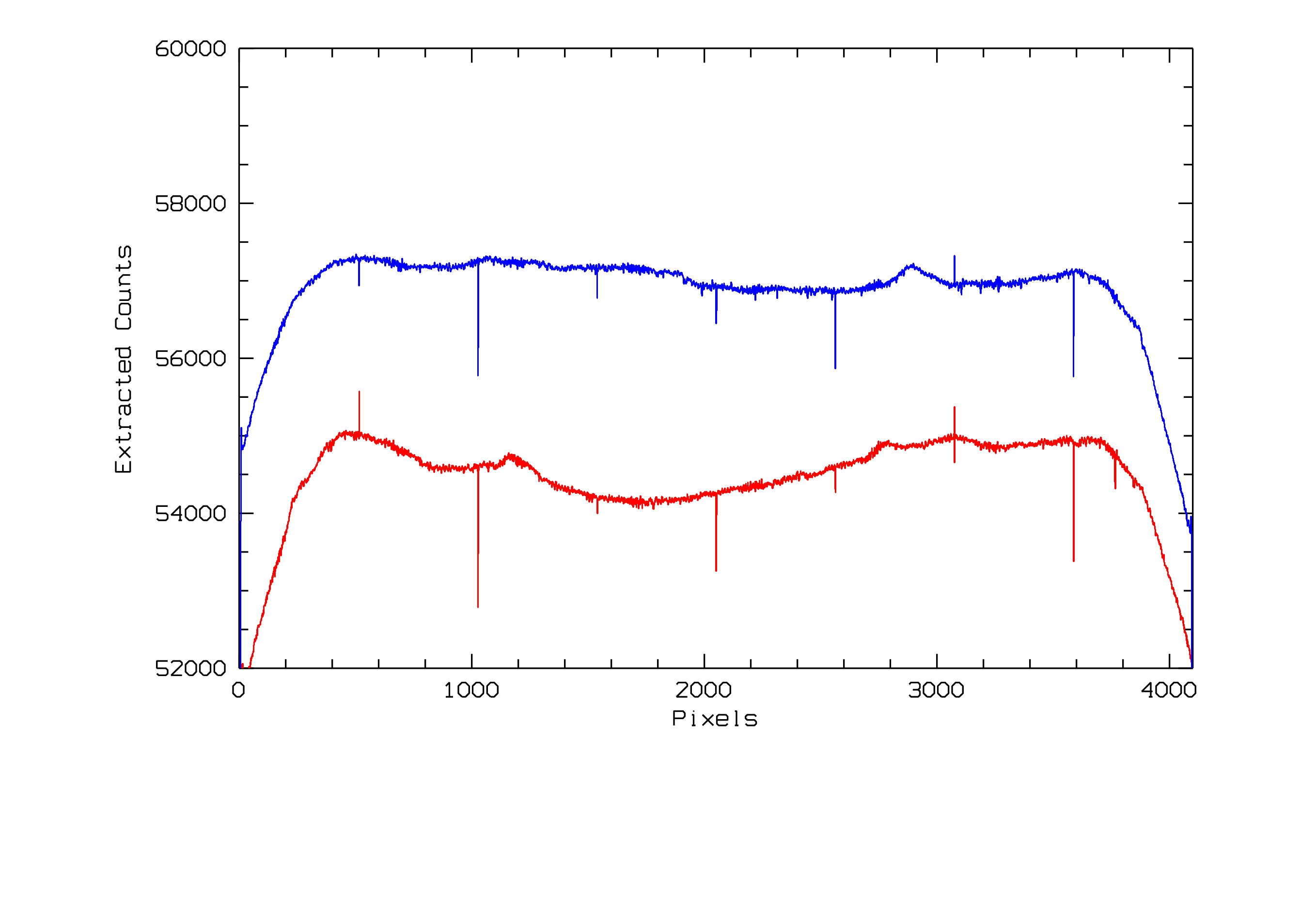}}
\caption{Sum up over columns of    technical LED flats revealing the  stitching for the  blue (top) and  red (bottom) HARPS CCDs  every 512 pixels.} 
\end{center}
\label{fig9}
\end{figure}


 Therefore,  the new positions provided by LP07 are particularly useful to minimize the effect of spectral distortions in HARPS spectra, but their use should not be   generalized to  other spectrographs.  We also performed a test by  comparing  a solar  spectrum calibrated with PE83   Th-Ar  lines  with the same spectrum but calibrated  with the  Th-Ar lines by LP07. In the spectral region 530-680 nm the two spectra reveal a difference  of  17  \ms in radial velocity.
The fact that   the reference spectrum of Molaro and Monai (2012) was calibrated with the PE83 line list  accounts for most of the difference  of $\approx$ 21 \ms between the expected velocity and the computed one reported in Table \ref{table:2}.

The  laboratory wavelength list  of LP07  was    modified to match  the lower resolving power of UVES by means of an  algorithm derived by  \citet{mur07}. These    are currently used in the UVES spectrograph. We emphasize  that since the modified positions are associated to HARPS distortions  there is no  reason to   export them  to a different spectrograph where such a distortions, even when present,  should have a different behaviour. A new set of  wavelengths for Th-Ar  based on this set of  LFC observations in the 475-580 nm spectral region is in preparation \citep{zao13}.

 \begin{figure}[]
\begin{center} 
\resizebox{!}{5.cm}{\includegraphics[clip=true,angle=0]{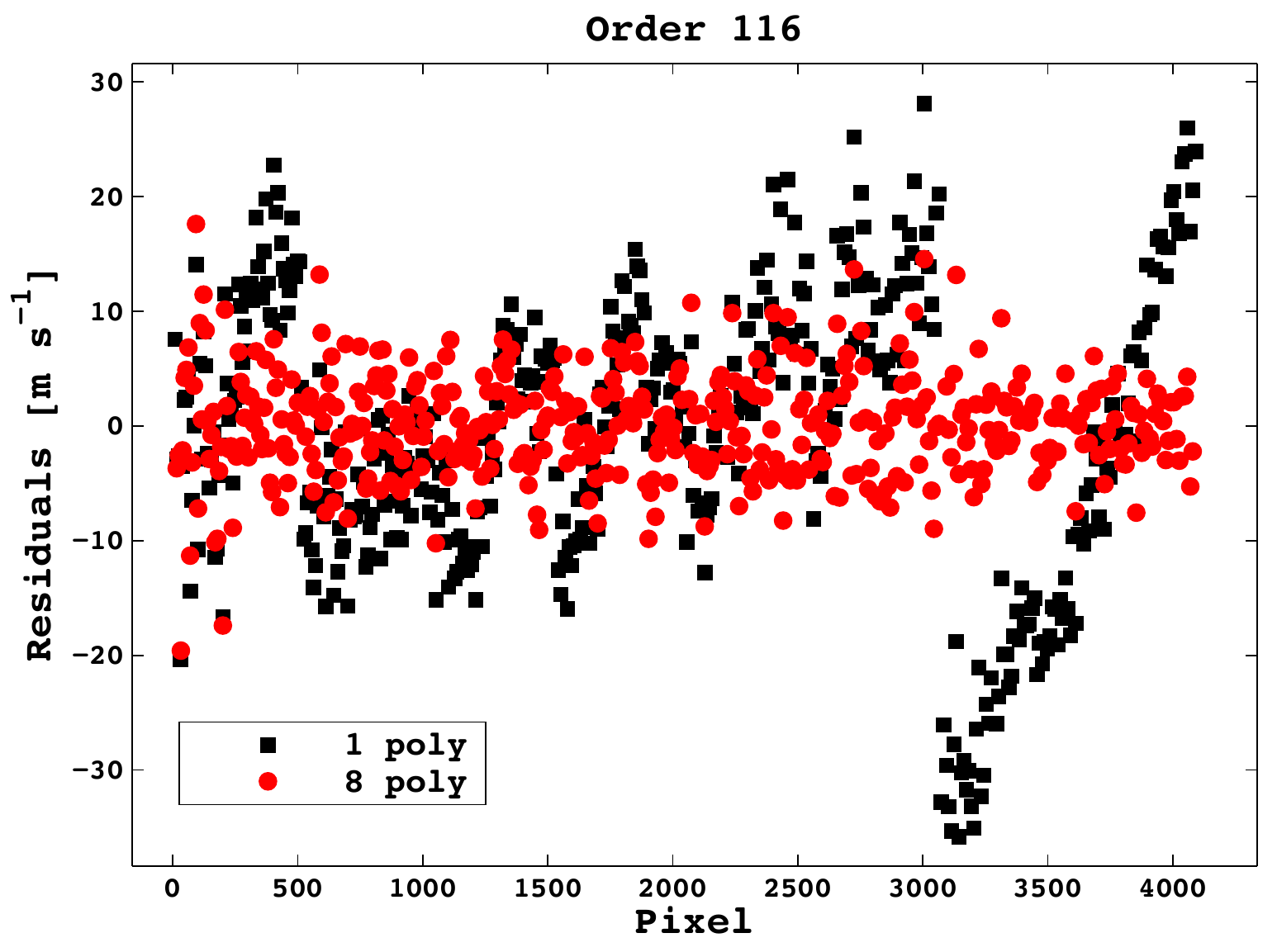}}
\caption{Residuals  of the two different calibrations  for Ord 119. The red points are those obtained after dividing the CCD by 8 different chunks of 512 pixels along the dispersion direction and fitted with a polynomial of 3rd degree. The black squares are those obtained by treating the  CCD as one made  by pixels of  equal size and same order polynomial. In the latter  enhancements in the residuals  show up in correspondence of the  CCD stitching borders.}
\end{center}
\label{fig10}
\end{figure}

\section{Summary and future prospects}

From the analysis of a high signal-to-noise  integrated solar spectrum   obtained from  Moon observations    and calibrated  by means of an LFC we  confirmed  the presence of wavelength distortions in the HARPS spectra first  revealed by \citet{wil10a} with the first LFC experiment. We also extended such an evidence to all  23   orders of the Echellogram covered by the LFC showing that the behavior is quite general and  should affect  the whole spectrum. We also showed that   the distortions in the different orders  have similar patterns but they  differ  among them depending on the number and position of Thorium  lines available in each order.     

By using the spectral region covered by the LFC we provided a new atlas of  solar line positions  with  about 600  entries in the range $
476-585$ nm.      The mean accuracy  of  individual  positions  of strong lines is of 12 $\pm$ 6 \ms and in many lines  can be as good as    $\le$ 5  \ms  and  is   free of  instrumental effects. 

Thus, the   atlas can be  used  to calibrate any  ground or space spectrographs  in the covered range down to this  precision.  As a whole,  it    provides  an  absolute reference of the solar photospheric radial velocity relative to the date of Nov 2010.  On   Nov 2010  the Sun  was   showing a reduced   activity of the solar cycle. The number of spots at that  epoch  is of 14.4,  a number which  corresponds to    a  low magnetic  activity.


The connection between  radial velocity  variations with  the cycle of magnetic activity along the solar cycle is far from  being settled. 
For instance several radial  velocity shifts of $\alpha$  Cen B related to stellar activity were  observed and corrected for in order to reveal the presence of one of the smallest  exoplanets \citep{dum12}.   
\citet{jim86}  measured changes in the  radial velocity of the Sun  in the period from 1976 and 1985 with  a resonant scattering spectrometer and found  variations of  16 \ms in correspondence of   the peak cycle and 
\citet{dem94}  measured    a peak-to-peak modulation
 of 30 \ms\ over  the 11-year  solar activity period with the positions 
redshifted in correspondence  to the maximum of the activity.   However, 
\citet{mcmil93} did  not reveal any drift  within 4 \ms\ in the solar line position  
from a long data series  spanning  from 1987 to 1992. 
\citet{lag10, lag11} computed the radial velocity variations caused by the spots observed on the Sun's  surface in cycle 23 (between 1993 and 2003) and included them in a synthetic model of the solar spectrum. They found typical amplitudes of up to a few \ms depending on the  activity level and on the temperature of the spot, but they were lower than those  observed by \citet{jim86} and \citet{dem94}. The periodograms contain peaks at periods that are sometimes very different from the solar rotation period and therefore require accurate control  to be revealed.  Contrary to   spots,   plages  suppress convection  in the active regions,  leading always to a redshift of the whole solar spectrum. \citet{meu10} considered the presence of plages and found a  long-term amplitude of 8 \ms while the short term radial velocity amplitude is  of  the order of  \ms or down to 0.4 \ms during the low-activity period. \citet{gra09} pointed out that the shift is line-dependent and that strong and weak lines should give  different shifts, which would be interesting to measure.  Since the various experiments used different lines at different wavelengths, this could, at least partially, explain the discrepancies.

 The solar spectrophotometer Mark-I located at the Observatorio del Teide  provides radial velocity  measurements of the Sun-as-a-star spanning the last 36 years and showed a clear modulation of $\pm$ 5 \ms in phase with the magnetic activity cycle and clearly anti-correlated with the solar spot number \citep{palle12}.   However, no dependence on the solar cycle is visible in the Global Oscillation at Low Frequencies (GOLF) observations   on board of the Solar heliospheric Observatory (SOHO) over 8 years of observations \citep{gar05}.  These activity-related movements  are expected at the level  of few \ms and  can be established with     HARPS observations of asteroids when calibrated with a laser frequency   comb.

The present list of absolute solar line positions could be also used as a reference of theoretical models of the solar photosphere. 
The 3D radiative-hydrodynamical convection models for FeI and FeII solar lines can be compared directly with the observed line positions. According to  \citet{asp2000}   the comparison of the absolute line shifts of their
models with the observed line positions  from \citet{AllendePrieto98b}  reveals
excellent agreement for weak FeI lines (EW$<$   0.006 nm  ) with a difference of  0 $\pm$  53 \ms  which degrades   to   51 $\pm$81 \ms  when including also   lines  with larger equivalent widths  (EW$< $ 0.01 nm ).
For  FeII lines the agreement is   slightly worse   with a mean difference of -64  $\pm$ 85 \ms .    3D radiative  hydrodynamical
models  of the solar photosphere computed with the CO5BOLD code have also been obtained and described in \citet{caf07}. For instance these models  for the FeI $\lambda\lambda$ 552.5544 nm  line  provide  a convective blue  shift of -371 \ms \citep{caf2010}.   In
the present atlas  the observed line position for this FeI line is  at  552.554816 nm and therefore it results redshifted
by +226 \ms   with respect to the laboratory wavelength. Adopting  a solar gravitational redshift of +633.5  \ms,  as measured at the Earth,    the intrinsic motion of the convective region of the  formation  of the FeI $\lambda\lambda$ 552.5544 nm  line is    blue-shifted by -407.5  \ms.  Thus the difference between the
observed central wavelength and the predicted one is of 36.5 \ms.     This example  shows how an improved solar line position could be useful in constraining theoretical models for the solar photosphere and a detailed comparison will be the subject of a dedicated work  \citep{gon13}.

 \begin{acknowledgements}
We are grateful to Peter Stetston and Elena Pancino for making available and providing assistance  with  DAOSPEC.   We thank  Simonetta Fabrizio and Gabriella Schiulaz for  reviewing the text.  We thank an anonymous referee for several suggestions which improved the paper.
R.R., M.E. and J.I.G.H. acknowledge financial support from the 
Spanish Ministry project MINECO AYA2011-26244, and
J.I.G.H. also from the Spanish Ministry of Economy and Competitiveness
(MINECO) under the 2011 Severo Ochoa Program MINECO SEV-2011-0187. L. P. is Visiting Researcher (PVE) of the CNPq Brazilian Agency, at the
Federal University of Rio Grande do Norte, Brazil.
\end{acknowledgements}

\bibliographystyle{aa}

\begin{thebibliography}{}


\bibitem[Allende Prieto and Garcia Lopez(1998a)]{AllendePrieto98a} Allende Prieto  C., \& Garcia Lopez  R. J. 1998a, A\&AS, 129, 41

\bibitem[Allende Prieto and  Garcia Lopez (1998b)]{AllendePrieto98b} Allende Prieto, C., \&  Garcia Lopez R. J. 1998b, A\&AS, 131, 431
\bibitem[Asplund et al (2000)]{asp2000} Asplund, M., Nordlund, A., Trampedach, R., Allende Prieto, C., \& Stein, R. F.
2000, A\&A, 359, 729

\bibitem[Brault and Neckel (1987)]{brau87} Brault, J., \& Neckel, H. 1987, Spectral Atlas of Solar Absolute Disk-Averaged and Disk-Center Intensity from 3290 to 12 510 A, unpublished (tape-copy from KIS IDL library) 
 

 \bibitem[Caffau and Ludwig (2007)]{caf07} Caffau E., and Ludwig, H. G., 2007 A\&A 467, L11
  \bibitem[Caffau (2010)]{caf2010} Caffau E., private comunication.
 \bibitem[Centuri{\'o}n et al. (2010)]{cen10}
Centuri{\'o}n, M. and Molaro, P. and Levshakov, S.  2010 Memorie SAIt, 80, 929, ArXiv 0910.4842





\bibitem[de Cuyper and Hensberge (1988)]{decuyper88} de Cuyper, J.-P., \& Hensberge, H. 1988, A\&AS, 129, 409

\bibitem[Deming and Plymate (1994)]{dem94} Deming, D., \& Plymate, C. 1994, ApJ, 426, 382

\bibitem[Dumusque et al. (2012)]{dum12} Dumusque, X., et al., 2012, Nature 491, 207

\bibitem[ Fei Zaho et al.  (2014)]{zao13}  Fei Zaho et al., 2014 in preparation.

\bibitem[Famaey et al. (2005)]{fam05} Famaey, B., Jorissen, A., Luri, X., et al., 2005, A\&A, 430, 165
  
\bibitem[Giacchetti et al.  (1970)]{gia70} Giacchetti, A., Stanley R., W., and Zalubas, R.,  1970, J. Opt. Soc. Am. 60, 474 

\bibitem[Garc\'ia et al. (2005)]{gar05} Garc\'ia R.A. 2005 A\&A 385, 395 

\bibitem[Gonz\'alez Hern\'andez et al. (2014)]{gon13}   Gonz\'alez Hern\'andez, J., I., et al  in preparation.
  
\bibitem[Gray and  Livingstone (1997)]{gra97} Gray, D. F., \& Livingston, W. C. 1997, \apj, 474, 802
 
\bibitem[Gray et al. (2000)]{gra00} Gray, D. F., Tycner, C., \& Brown, K. 2000, PASP, 112, 328
\bibitem[Gray et al. (2009)]{gra09} Gray, D. F., 2009 \apj,  697:1032

\bibitem[Hunten (1970)]{hun70} Hunten, D. M. 1970 \apj 159, 1107 


 

\bibitem[Kurucz et al. (1984)]{kurucz84} Kurucz, R. L., Furenlid, I. J.,  \& Testerman, L. 1984, 
{\it NOAO Atlas No.~1, The Solar Flux Atlas from 296 to 1300 nm. Sunspot}, 
(NM: National Solar Observatory)

\bibitem[Kurucz (2006)]{kurucz06} Kurucz, R. L., arXiv:astro-ph/0605029

\bibitem[Jim\'enez et al. (1986)]{jim86} Jim\'enez, A., Palle, P.L., Regulo, C. et al. 1986, AdSpR 6, 89



\bibitem[Lagrange et al. (2010)]{lag10} Lagrange A. -M., Desort M., Meunier N. 2010 A\&A 512, A38
\bibitem[Lagrange et al. (2011)]{lag11} Lagrange A. -M., Desort M., Meunier N. 2011  A\&A  528, L9 



\bibitem[Lindegren and Dravins (2003)] {lin03} Lindegren, L., \& Dravins, D. 2003, A\&A, 401, 1185

 
 \bibitem[Livingston et al. (1999)]  {liv99}
 	Livingston, W.,  Wallace, L., Huang, Y., Moise, E 1999 ASPC 183, 494
\bibitem[Lovis and Pepe (2007)]{lovis07} Lovis, C., \& Pepe F.  2007, A\&A, 468, 1115


\bibitem[Mayor et al. (2003)]{may03}
Mayor, M., Pepe, F., Queloz, D., Bouchy, F., Rupprecht, G., Lo Curto, G., Avila, G., Benz, W., Bertaux, J.-L., Bonfils, X., Dall, Th., Dekker, H., Delabre, B., Eckert, W., Fleury, M., Gilliotte, A., Gojak, D., Guzman, J. C., Kohler, D., Lizon, J.-L., Longinotti, A., Lovis, C., Megevand, D., Pasquini, L., Reyes, J., Sivan, J.-P., Sosnowska, D., Soto, R., Udry, S., van Kesteren, A., Weber, L., Weilenmann, U. 2003 Msngr 114, 20


\bibitem[Mayor et al. (2009)]{may09} Mayor, M.,  Bonfils, X.;, Forveille, T.,  Delfosse, X.,  Udry, S,  Bertaux, J.-L., Beust, H., Bouchy, F., Lovis, C., Pepe, F., Perrier, C., Queloz, D., Santos, N.C. 2009 A\&A 507, 487
 
\bibitem[McMillan et al. (1993)]{mcmil93} McMillan, R. S., Moore, T. L., Perry, M. L., \& Smith, P.H. 
1993, ApJ, 403, 801

\bibitem[Meggers (1957)]{meg57} Meggers, W.F.,  1957 in Oosterhoff P.T., ed Trans. IAU Vol. 9 proceedings of the Ninth General Assembly, Cambridge Univ. press, cambridge, p225 


\bibitem[Meunier et al. (2010)]{meu10} Meunier N.,  Desort M., Lagrange A. -M.,  2010 A\&A 512, A39


\bibitem[Mitchell et al. (1991)]{} Mitchell, W. E., Jr., and Livingston, W., C., 1991, ApJ, 372, 336

\bibitem[Molaro et al. (2008)]{mol08a}
Molaro, P., Levshakov, S.~A., Monai, S., et al. 2008, \aap, 481, 559

\bibitem[Molaro et al. (2008b)]{mol08b}
Molaro, P. , Reimers, D. ,Agafonova, I.~I. and Levshakov, S.~A.
2008, EPJ, 163, 173

 \bibitem[Molaro and  Monai  (2012)]{mol12} Molaro, P. , Monai, S., 2012 A\&A 544,125
 \bibitem[Molaro and Centurion (2011)]{mol11} Molaro, P. , Centurion, M., 2011 A\&A 525, A74

 \bibitem[Murphy et al.  (2007)]{mur07} Murphy, M. T.; Udem, Th.; Holzwarth, R.; Sizmann, A.; Pasquini, L.; Araujo-Hauck, C.; Dekker, H.; D'Odorico, S.; Fischer, M.; H\"ansch, T. W.; Manescau, A. 2007 MNRAS 380, 839.

\bibitem[Murphy et al. (2007b)]{murphy2007} Murphy, M. T., Tzanavaris, P., Webb, J. K., \&  Lovis, C.
2007,  MNRAS, 378, 221

\bibitem[Murphy et al. (2006)]{murphy2006} Murphy, M. T.,  Webb, J. K., \&  Flambaum V.V. 
2006, arXiv: 0612407

 
 \bibitem[Nidever et al. (2002)]{nid02} Nidever, D. L., Marcy, G. W., Butler, R.P., et al., 2002, ApJS, 141, 503
 
 \bibitem[Nordstr{\"o}m et al. (2004)]{nor04} Nordstr{\"o}m, B., Mayor, M., Andersen, J., et al., 2004, A\&A, 418, 989
 \bibitem[Palmer (1983)]{PE83} Palmer, B. A., Engleman, R.  1983, LA, Los Alamos: National Laboratory, |c.
 
 \bibitem[Palle et al.  (2013)]{palle12} Pall\'e, P.L., and Cort\'es T. Roca 2013 Highlights of Spanish Astrophysics VII, Proceedings of the X Scientific Meeting of the Spanish Astronomical Society (SEA), held in Valencia, July 9 - 13, 2012, Eds.: J.C. Guirado, L.M. Lara, V. Quilis, and J. Gorgas., pp.750-761
 
\bibitem[Pepe et al. (2005)]{pepe05} Pepe, F., Mayor, M., Queloz, D., et al.  2005, The Messenger, 120, 22


\bibitem[Phillips et al. (2012)]{phi12}  Phillips, D., Glenday, A.,  Li, C.,  Cramer, C.,   Furesz, G. Chang, G.,  Benedick, A.,  Chen, L.,   K\"artner,  K., Korzennik, S.,  Sasselov, D.,  Szentgyorgyi, A., and  Walsworth,  R., 2012, Opt. Express  20, 13711-13726. 

\bibitem[Redman et al. (2013)]{red13} Redman, S. L., Nave, G., Sansonetti C., J. 2013 arXiv:1308.5229v1 
\bibitem[Steinmetz et al. (2008)]{stei08} Steinmetz, T., 
Wilken, T., Araujo-Hauck, C., et al.\ 2008, Science, 321, 1335 

 \bibitem[Stetston and Pancino (2008)]{ste08} Stetston, P. B.  Pancino, E. 2008 PASP 120, 1332  
 \bibitem[Thomas et al. (2005)]{tho05} Thomas, P.C., Parker J. Wm., McFadden, L.A., Russel C. T., Stern S. A. Sykes M. V. and Young E. F., 2005,  Nature 437, 224 
       
  

\bibitem[Wallace et al. (2011)]{wal11} Wallace L., Hinkle, K., Livingston W.C., Davis, S.P., N. S. O. Technical Report N. 11-001
Published 2011, by National Solar Observatory, P. O. Box 26732, Tucson AZ 85726


\bibitem[Wilken et al. (2010b)]{wil10b} Wilken, T. et al., 2010 MNRAS 405, 16
\bibitem[Wilken et al. (2010a)]{wil10a} Wilken, T. et al., 2010 SPIE.7735E..28W
\bibitem[Wilken et al. (2012)]{wil12} Wilken, T et al., 2012 Nature 485, 611.
 
\bibitem[Ycas et al. (2012)]{yca12}   Ycas, G., Quinlan, F.,   Diddams, S. Osterman, S. Mahadevan, S. Redman, S., Terrien, R.,  Ramsey, L.,   Bender,C.,   Botzer, B.,  and  Sigurdsson,  S.,  2012 Opt. Express  20, 6631-6643. 

\end{thebibliography}


 \end{document}